\documentclass[journal=jpcbfk,layout=twocolumn,manuscript=article]{achemso}
\usepackage{multirow}
\usepackage{rotating}
\author{Marcello Sega}
\author{Christoph Dellago}
\email{marcello.sega@univie.ac.at}
\affiliation[University of Vienna]
{Faculty of Physics, University of Vienna, Boltzmanngasse 5, A-1090, Vienna, Austria} 

\title[Long-range dispersion effects on water/vapor model interfaces]
  {Long-range dispersion effects on the water/vapor interface simulated using the most common models}

\begin{document}

\begin{tocentry}
\includegraphics[width=\columnwidth]{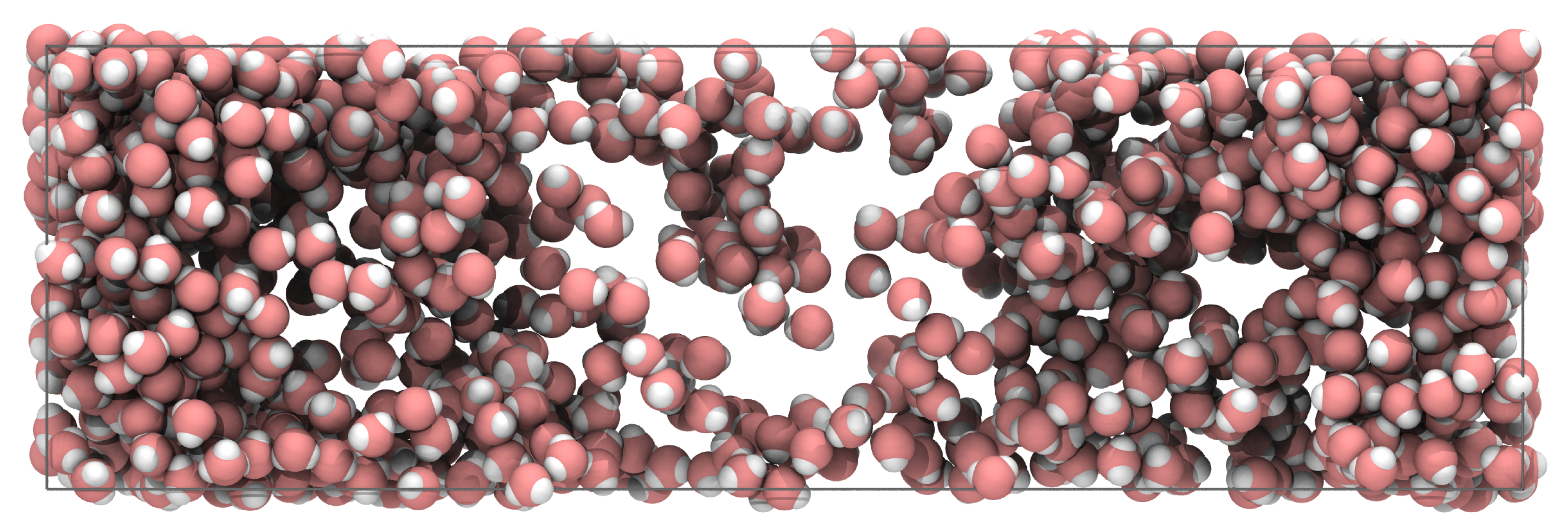}
\end{tocentry}

\begin{abstract}
The long-range contribution to dispersion forces is known to have a major impact on the properties of inhomogeneous fluids, and its correct treatment is increasingly recognized as being a necessary requirement to avoid cutoff-related artefacts. Although analytical corrections for quantities like the surface tension are known, these can not take into account the structural changes induced by the long-range contributions. Here, we analyze the interfacial properties of seven popular water models, comparing the results with the cut-off version of the dispersion potential. The differences in surface tension estimates are in all cases found to be less than 2 mN/m.
\end{abstract}

\section{Introduction}
Efficiency considerations have often been at the root of the practice, customary in computer simulations of molecular systems\cite{hansen90}, to truncate dispersion forces between pairs of particles separated by more than a given cut-off. In contrast to the electrostatic force, whose contribution from far particles can be comparable to that of close ones\cite{DeLeeuw1980}, dispersion forces are decaying much faster, providing often a justification for the truncation. The importance of taking into account the full dispersion forces, however, has been known since long in the field of crystallography, where it is essential for the precise estimation of binding energies (it is no wonder that perhaps the most celebrated method for the calculation of the full electrostatic energy, developed by Ewald\cite{ewald21a}, was devised precisely to compute the energy of ionic crystals). For homogeneous, isotropic liquids, the analytical tail corrections to the energy and (more importantly) pressure of dispersion forces are usually sufficient to remove artifacts due to truncation. Analytical corrections to the surface tension are also known~\cite{Blokhuis1995}, that apply to inhomogeneous systems in slab configuration. 

Explicit simulations of the liquid/vapor coexistence are usually performed at fixed volumes, and the correction to the surface tension cannot be used as a feedback to change the density of the fluid (as in the case of constant pressure simulations). In't~Veld, Ismail and Grest~\cite{IntVeld2007a} showed that taking into account the full dispersion forces has a dramatic impact on the density profile of the Lennard-Jones liquid/vapor interface. Similarly, Wennberg and coworkers demonstrated the importance of the long-range part of dispersion forces in membranes~\cite{wennberg13}. These effects, although less pronounced in the case of pure water~\cite{IntVeld2007a} because the dominant contribution to the cohesion forces is of electrostatic nature, follow a clear trend, observable in the binodal line, of increasing the density of the liquid phase. As we will show, the analytical corrections tend to overestimate the surface tension for temperatures $T$ lower than $\simeq 450~K$, and to underestimate it for larger ones, although always within about 2~mN/m.

In the following sections we will first discuss our methodological approach, including the simulation details and the calculation of the analytical tail contribution, and then we will present the results of our simulations on the binodal line and on the surface tension of several, commonly used water models. 

\section{Methods}
\begin{figure}[t]
\includegraphics[width=\columnwidth]{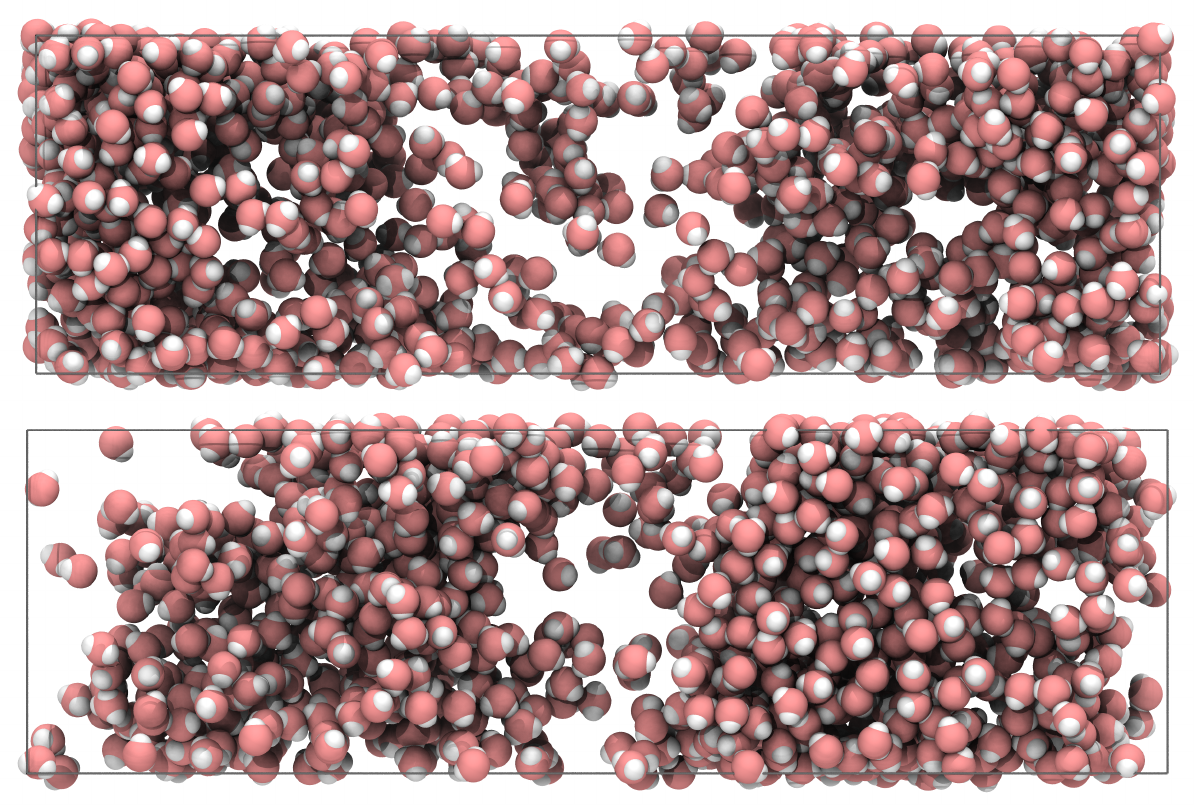}
  \caption{Snapshots of two equilibrium configurations of TIP4P water (550~K, no long-range dispersion forces). Upper panel: the system is clearly composed of one liquid phase and one vapor phase. Lower panel: the system separates into two separate liquid slabs.}
  \label{fig:snap}
\end{figure}

We simulated seven popular rigid water models, including three-, four-, and five-point ones, namely, SPC\cite{hermans84},SPC/E\cite{berendsen87a}, TIP3P\cite{Jorgensen1983}, TIP4P\cite{Jorgensen1983},TIP4P/2005\cite{Abascal2005},TIP5P\cite{mahoney00a} and, TIP5P-E\cite{rick2004reoptimization}. All simulations were performed using the GROMACS molecular dynamics simulation package, version 5.1\cite{Abraham2015}, and consisted of 1000 water molecules in a slab configuration in a rectangular simulation box of $5\times 5\times 15$~nm$^3$, with periodic boundary conditions applied in all directions. The simulations were performed in the canonical ensemble by integrating the equation of motions using the leapfrog algorithm (1~fs timestep), imposing the temperature equilibrium value by means of a Nos\'e--Hoover thermostat\cite{nose84a,hoover85a} (2~ps relaxation time), and keeping the molecules rigid using the SETTLE algorithm\cite{miyamoto92}. The electrostatic energy, force, and pressure were calculated using the smooth Particle Mesh Ewald method\cite{Essmann1995a} (sPME) with a real space mesh spacing of 0.15 nm, a $10^{-5}$ relative accuracy of the potential energy at the real space cutoff of 1.3 nm, and an interpolation scheme of order 4. In all simulated models the dispersion forces are taken into account by using the Lennard-Jones pair interaction. For each model and each chosen temperature $T$ (300, 350, 400, 450, 500 and 550~$K$) we performed one simulation by truncating dispersion forces at $r_c=1.3$~nm, and one by taking into account the contribution of periodic copies again using sPME\cite{Essmann1995a,Wennberg2015} , with a $10^{-3}$ relative accuracy at the real space cut-off of 1.3~nm (the same as in Ref.~\cite{Vega07}). 
After an equilibration of 2~ns, an equilibrium trajectory of 50~ns was generated, saving configurations to disk at 1~ps intervals, and pressure every 0.1~ps.
The density profiles were calculated as $\rho(z) = \left\langle\sum_i\delta(z-z_i+z_\textrm{cm})\right\rangle$, where the sum is extended over all molecules, $z_\textrm{cm}$ is the position of the center of mass of the system, and the angular 
brackets represent the ensemble average. Prior calculation of the profile, the liquid 
phase was determined in every frame using a cluster search based on a distance criterion: two molecules 
that dist less than 0.35~nm are considered to be in the same cluster, and the largest 
cluster in the system identifies the liquid phase. The liquid phase slab was then shifted by a suitable amount along the macroscopic interface normal, in order not to cross 
the box boundaries, thus removing the ambiguity in the definition of the center of mass of 
a periodic system.

\begin{figure}[t]
\includegraphics[width=\columnwidth, trim={0 0 0 10},clip]{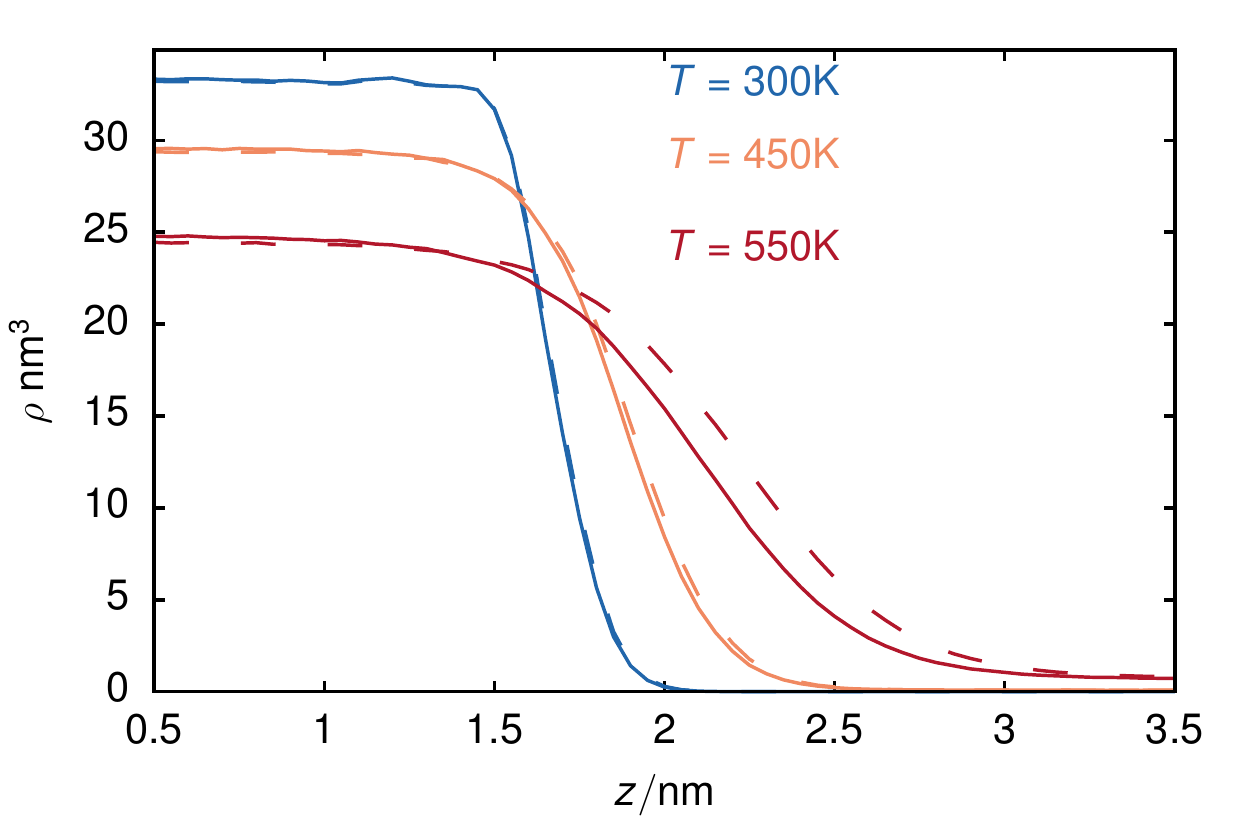}
  \caption{Density profile in the interfacial region for the TIP4P-2005 model, with (full lines) and without (dashed lines) long range Lennard-Jones contribution. The origin of the $z$-axis is located at the center of mass of the system.}
  \label{fig:dens}
\end{figure}

The analytical correction to the surface tension~\cite{Vega07,Blokhuis1995} can be obtained by calculating the integral
\begin{eqnarray}
\gamma_\mathrm{tail} = 12\pi \epsilon \sigma^6 (\rho_L -\rho_V)^2\int_0^1 ds \int_{r_c}^\infty dr \times\nonumber\\\times \coth(rs/d) (3s^3-s) r^{-3}.\label{tail}
\end{eqnarray}
The densities of the liquid ($\rho_L$) and of the vapor ($\rho_V$) phases as well as the interface thickness $d$ can be estimated by performing a Marquardt-Levenberg least square fit of the sampled density profiles (see Fig.~\ref{fig:dens} for an example) to the function 
\begin{equation}
\rho(z) = \frac{1}{2}(\rho_L+\rho_V) - \frac{1}{2}(\rho_L-\rho_V) \tanh[(z-z_0)/d],\label{profile}
\end{equation}
where $z$ is the position along the interface normal, and $z_0$ identifies the location of the middle of the interface. In order to perform the integral, Eq.~(\ref{tail}), we used the  general purpose adaptive quadrature for infinite intervals (QAGI) of the QUADPACK library~\cite{piessens2012quadpack}, as simpler integration schemes showed a marked dependence on the choice of the upper limit of integration. 

At the highest investigated temperature of 550~K, some of the systems, especially after long simulation times, separate occasionally into two or more liquid slabs, preventing to perform a meaningful fit of the density profiles with Eq.~(\ref{profile}) and, at the same time, preventing to compute the surface tension with the help of the expression for planar interfaces 
\begin{equation}
\gamma = \frac{L}{2} ( p_N - p_T),
\end{equation}
where $L$ is the length of the box edge parallel to the surface normal, and $p_N$ and $p_T$ are the normal and  lateral pressure components, respectively. In these cases, we did not determine the data for surface tension and densities of the liquid and vapor phases. In Fig.\ref{fig:snap} we report two snapshots of the TIP4P water/vapor interface  at $T$=550~K, showing a case where the system is partitioned into two distinct liquid and vapor phases (top) and one case where the system separated into two liquid droplets surrounded by vapor (bottom). 

\section{Results and Discussion}
\begin{figure}[t]
 \includegraphics[width=\columnwidth, trim={10 0 10  30},clip]{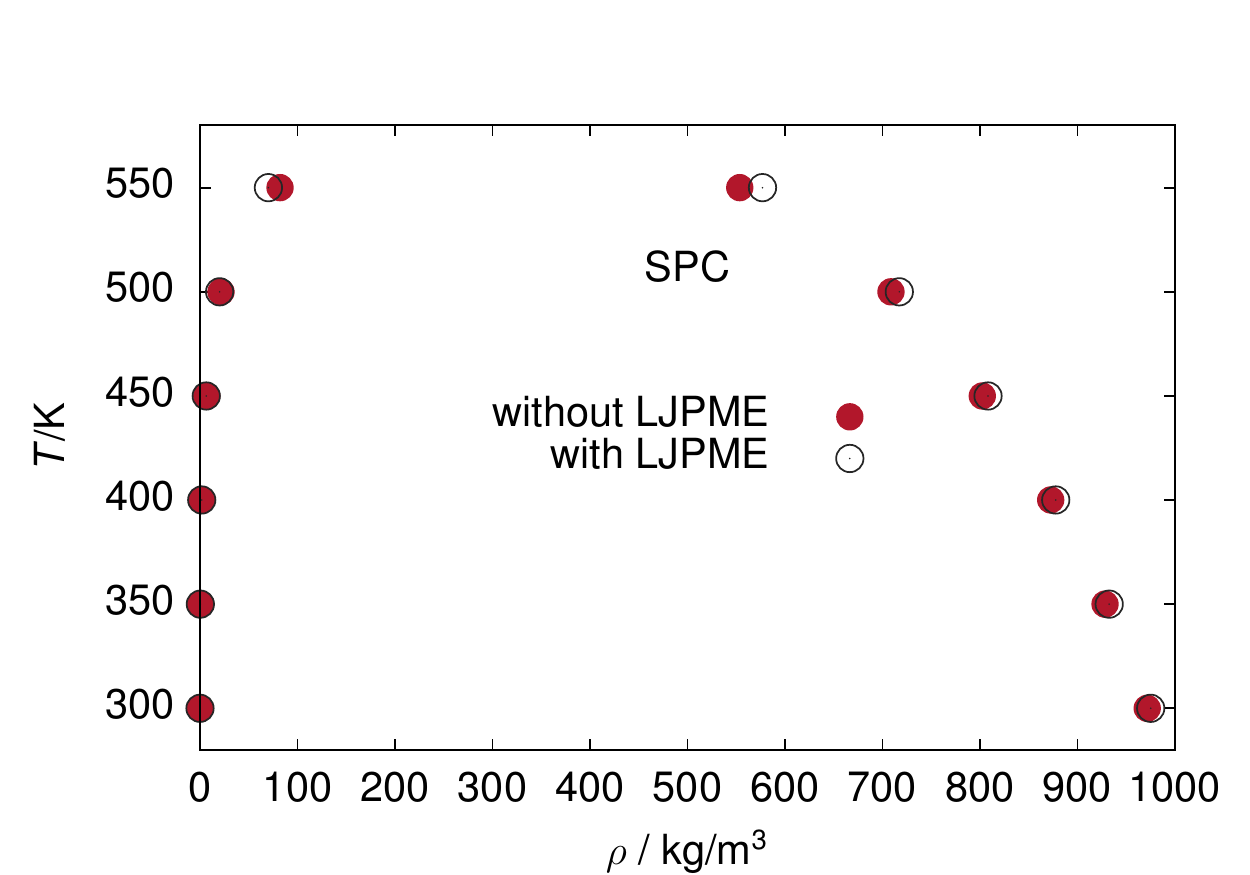}
  \caption{Density of the coexisting liquid and vapor phases for the SPC water model, as a function of the temperature. Full symbols: results from the simulations with truncated dispersion forces; open symbols: results from the simulations with full long-range dispersion forces. The error bars are much smaller than the symbols.}
  \label{fig:rhoT}
\end{figure}
In Tab.~\ref{table1} we report the average densities of the liquid ($\rho_L$) and of the vapor ($\rho_V$) phases for  different temperatures, both for the case when dispersion forces are treated with sPME, and when they are truncated at $r_c=1.3$~nm. The values obtained from the simulation with truncated forces are reported as differences from the sPME results ($\Delta \rho_L$ and $\Delta \rho_V$). The same convention is used for the other reported quantities (interfacial width $d$, surface tension $\gamma$). With the exception of the  TIP4P-2005 model, the inclusion of  long-range dispersion forces with sPME shifted the equilibrium density of the liquid phase systematically to higher values, and the density of the vapor phase to lower values.

\begin{figure}[t]
\includegraphics[width=\columnwidth, trim={10 80 10 10},clip]{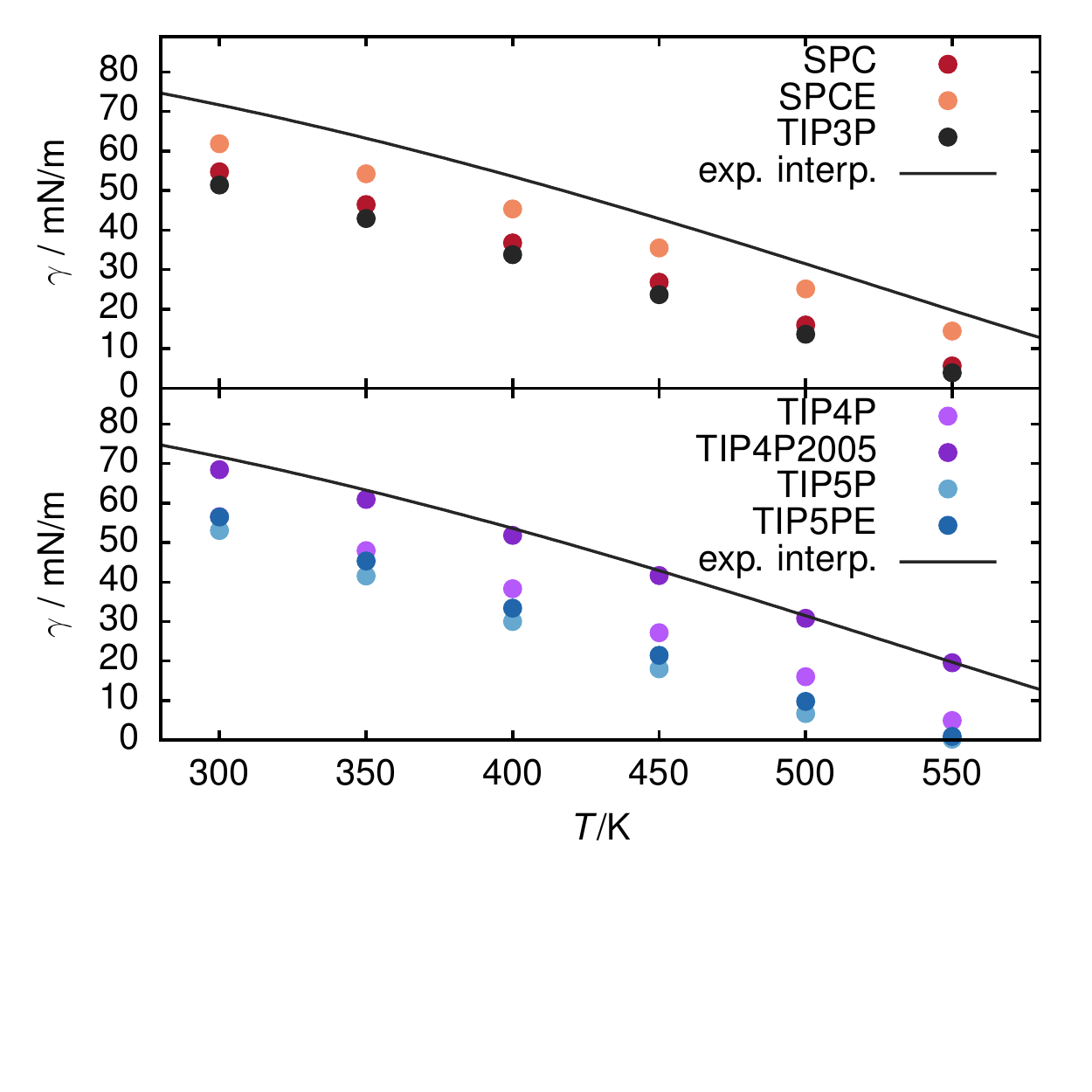}
  \caption{Values of the surface tension as a function of the temperature for the seven simulated water models, when long-range contribution of the dispersion forces are taken into account.}
  \label{fig:gamma}
\end{figure}

This effect is the more pronounced, the higher the temperature, and can be easily appreciated by plotting the densities along the coexistence curve, as a function of the temperature (see Fig.~\ref{fig:rhoT} for the SPC model and the supplementary material for the other systems). 

The fact that the density of the liquid phase increases is a consequence of the stronger cohesive forces arising by inclusion of long-range contributions. As a consequence, the difference $ \rho_L - \rho_V$ increases, and this implies a shift of the critical temperature to higher values and, therefore, of a higher surface tension, as we will see later. In retrospect, such a change in density could have been expected, as the analytical tail corrections for the surface tension are known to be positive~\cite{Vega07}. 

Another effect of the increased cohesive force is the change in interfacial width, which also decreases systematically when the long-range dispersion forces are taken into account. A general trend appears also in this case, with the interfacial width at higher temperature being characterized by larger differences, although in absolute terms the changes are rather small, in most cases being below one Angstrom, and at most half of a nanometer (see Tab.~\ref{table1}).

The case of TIP4P-2005, as already mentioned, is qualitatively different from the other models, in that not only the density of the 
liquid increases upon introduction of the long-range dispersion forces, but also that of the vapor. This fact does not 
contradict the increase in surface tension, as the difference between liquid and vapor densities is still larger due to the truncated 
dispersion force cases, but we have found no explanation for this opposite trend. 

The surface tension for the seven different water models, simulated including the long-range dispersion forces, is reported in Fig.~\ref{fig:gamma}, together with the continuous line that interpolates experimental results using the formula,\cite{Vargaftik1983} 
\begin{equation}
\gamma= B \tau^\mu (1+b \tau ) ,
\end{equation}
where $\tau =1-T/T_c $ , $ T_c=647.096$~K, $B=235.8$~mN/m, $b=-0.625$ and $\mu=1.256$.

\begin{table*}
\begin{small}
\begin{tabular}{ccccccccccc}
model & T & $\rho_L$ & $\Delta\rho_L$ & $\rho_V$ & $\Delta \rho_V$ & $d$ & $\Delta d$ & $\gamma$ & $\Delta \gamma$ & $\gamma_\textrm{tail} $\\
\hline\hline
    &300  & 975.2(2)   & 3.5   & 0.024(1)   & -0.007     & 1.644(3)   & -0.016   & 54.8(1) & 3.1   & 4.3\\     
    &350  & 932.6(2)   & 4.1   & 0.297(4)   & -0.016     & 2.062(3)   & -0.023   & 46.5(1) & 3.1   & 3.9\\ 
SPC &400  & 877.9(2)   & 5.3   & 1.721(10)   & -0.042    & 2.592(4)   & -0.025   & 36.8(2) & 2.5   & 3.3\\ 
    &450  & 808.4(2)   & 5.9   & 6.328(19)   & -0.624    & 3.338(5)   & -0.047   & 26.8(2) & 2.3   & 2.6\\ 
    &500  & 717.4(2)   & 8.6   & 20.10(4)   & -1.520     & 4.666(5)   & -0.090   & 16.0(2) & 2.1   & 1.8\\ 
    &550  & 576.9(2)   & 23.3  & 70.22(27)   & -11.9     & 7.789(14)   & -0.578  & 5.6(1)  & 1.3   & 0.6\\ 
\hline
   &300  & 996.8(2)   & 3.2     & 0.003(0)   & -0.004     & 1.473(4)   & -0.003  & 61.8(1)  &  2.8 & 4.5 \\
   &350  & 964.0(2)   & 3.5     & 0.103(2)   & -0.006     & 1.837(3)   & -0.007  & 54.3(2)  &  3.3 & 4.2 \\
SPC/E&400  & 919.1(2)   & 4.1   & 0.686(7)   & 0.043      & 2.269(3)   & -0.015  & 45.4(2)  &  2.7 & 3.7 \\
   &450  & 863.0(2)   & 5.1     & 2.700(13)  & -0.130     & 2.814(4)   & -0.024  & 35.5(2)  &  2.4 & 3.1 \\
   &500  & 792.9(2)   & 7.4     & 8.784(23)  & -0.480     & 3.620(4)   & -0.064  & 25.1(2)  &  1.9 & 2.4 \\
   &550  & 698.8(2)   & 10.2    & 25.61(4)   & -1.958     & 5.132(5)   & -0.154  & 14.5(2)  &  1.8 & 1.6 \\
\hline
     &300     & 983.1(2)   & 3.8    & 0.033(1)   & 0.003      & 1.733(3)   & -0.012  & 51.4(1)  & 3.4  & 4.6 \\
     &350     & 933.1(2)   & 4.2    & 0.359(4)   & -0.040     & 2.183(4)   & -0.012  & 43.0(1)  & 2.7  & 4.0 \\
TIP3P&400   & 871.5(2)     & 5.4    & 1.875(9)   & -0.090     & 2.751(4)   & -0.010  & 33.8(1)  & 2.5  & 3.4 \\
     &450     & 795.0(2)   & 7.5    & 7.116(19)   & -0.197    & 3.560(5)   & -0.060  & 23.7(1)  & 2.1  & 2.6 \\
     &500     & 694.2(2)   & 11.3   & 21.99(4)   & -2.064     & 5.158(6)   & -0.038  & 13.7(1)  & 2.0  & 1.6 \\
\hline
     & 300  & 991.7(2)   & 3.7      & 0.031(1)   & -0.003   & 1.609(3)   & -0.010   & 56.5(1)  &  3.1 & 4.6 \\
     & 350  & 953.3(2)   & 3.9      & 0.348(4)   & -0.046   & 2.024(3)   & -0.013   & 48.0(1)  &  3.3 & 4.2 \\
TIP4P& 400  & 899.2(2)   & 4.6      & 2.049(11)   & -0.297  & 2.553(3)   & -0.021   & 38.3(2)  &  3.3 & 3.6 \\
     & 450  & 828.8(2)   & 6.9      & 7.577(22)   & -0.483  & 3.326(4)   & -0.056   & 27.2(2)  &  2.3 & 2.8 \\
     & 500  & 734.7(3)   & 10.1     & 24.289(36)   & -1.033 & 4.724(5)   & -0.136   & 16.0(2)  &  2.1 & 1.9 \\
     & 550  & 582.4(3)   & -        & 82.47(28)   & -       & 8.316(17)  & -        & 4.9(1)   &  -   & -   \\
\hline
     & 300  & 995.3(3)   & 2.2    & 0.005(1)   & -0.080    & 1.407(4)   & -0.011  & 68.4(2)  &  3.8   & 4.6  \\
     & 350  & 971.2(2)   & 3.8    & 0.072(2)   & 0.072     & 1.734(2)   & -0.013  & 60.9(2)  &  3.8   & 4.3  \\
TIP4P/&400  & 932.7(2)   & 4.7    & 0.570(6)   & 0.570     & 2.130(3)   & -0.012  & 51.8(2)  &  3.4   & 3.8  \\
2005 & 450  & 882.9(2)   & 6.0    & 2.320(12)   & 0.686    & 2.624(4)   & -0.015  & 41.6(2)  &  3.1   & 3.3  \\
     & 500  & 820.0(2)   & 6.8    & 7.490(21)   & 0.446    & 3.325(4)   & -0.030  & 30.8(2)  &  2.7   & 2.7  \\
     & 550  & 739.7(2)   & 9.1    & 20.88(3)   & -0.202    & 4.484(4)   & -0.121  & 19.6(2)  &  2.5   & 1.9  \\
\hline
     & 300 & 981.8(2)   & 3.9     & 0.078(2)   & -0.006    & 1.750(3)   & -0.016  & 53.0(2)  & 2.9   & 4.6  \\
     & 350 & 936.7(2)   & 4.5    & 0.730(6)   & 0.004      & 2.289(4)   & -0.022  & 41.5(2)  & 2.9   & 4.1  \\
TIP5P& 400 & 863.7(2)   & 6.3    & 3.819(15)   & -0.094    & 3.030(5)   & -0.025  & 30.0(1)  & 2.4   & 3.3  \\
     & 450 & 763.0(2)   & 8.9     & 13.978(28)  & -0.897   & 4.241(5)   & -0.082  & 18.0(1)  & 2.0   & 2.2  \\
     & 500 & 614.0(3)   & 19.8    & 52.46(18)   & -7.128   & 7.165(10)  & -0.456  & 6.7(1)   & 1.2   & 0.9  \\
\hline
     & 300 & 1000.1(2)  & 4.0    & 0.051(2)   & -0.012     & 1.712(3)   & -0.006 & 56.5(2)  & 3.9  & 4.5\\
     & 350 & 956.8(2)   & 4.3    & 0.597(6)   & -0.028     & 2.215(4)   & -0.023 & 45.3(2)  & 3.3  & 4.0\\
TIP5PE&400 & 887.6(2)   & 6.0    & 3.075(13)   & -0.280    & 2.889(5)   & -0.022 & 33.4(1)  & 2.7  & 3.3\\
     & 450 & 794.8(2)   & 8.5    & 11.510(26)   & -0.359   & 3.957(5)   & -0.089 & 21.4(1)  & 2.1  & 2.4\\
     & 500 & 664.4(3)   & 15.6   & 38.033(61)   & -5.736   & 6.269(7)   & -0.212 & 9.8(1)   & 1.8  & 1.2\\
\hline
\end{tabular}
\end{small}
\caption{Various properties for the different models simulated with inclusion of long-range dispersion forces and the difference with the same properties calculated with cutoff (analytical corrections not included). Temperatures are expressed in K, densities in kg/m$^3$, distances in nm, and surface tensions in mN/m. }
\label{table1}
\end{table*}

\begin{figure}[t]
\includegraphics[width=\columnwidth,trim={0 20 25 10},clip]{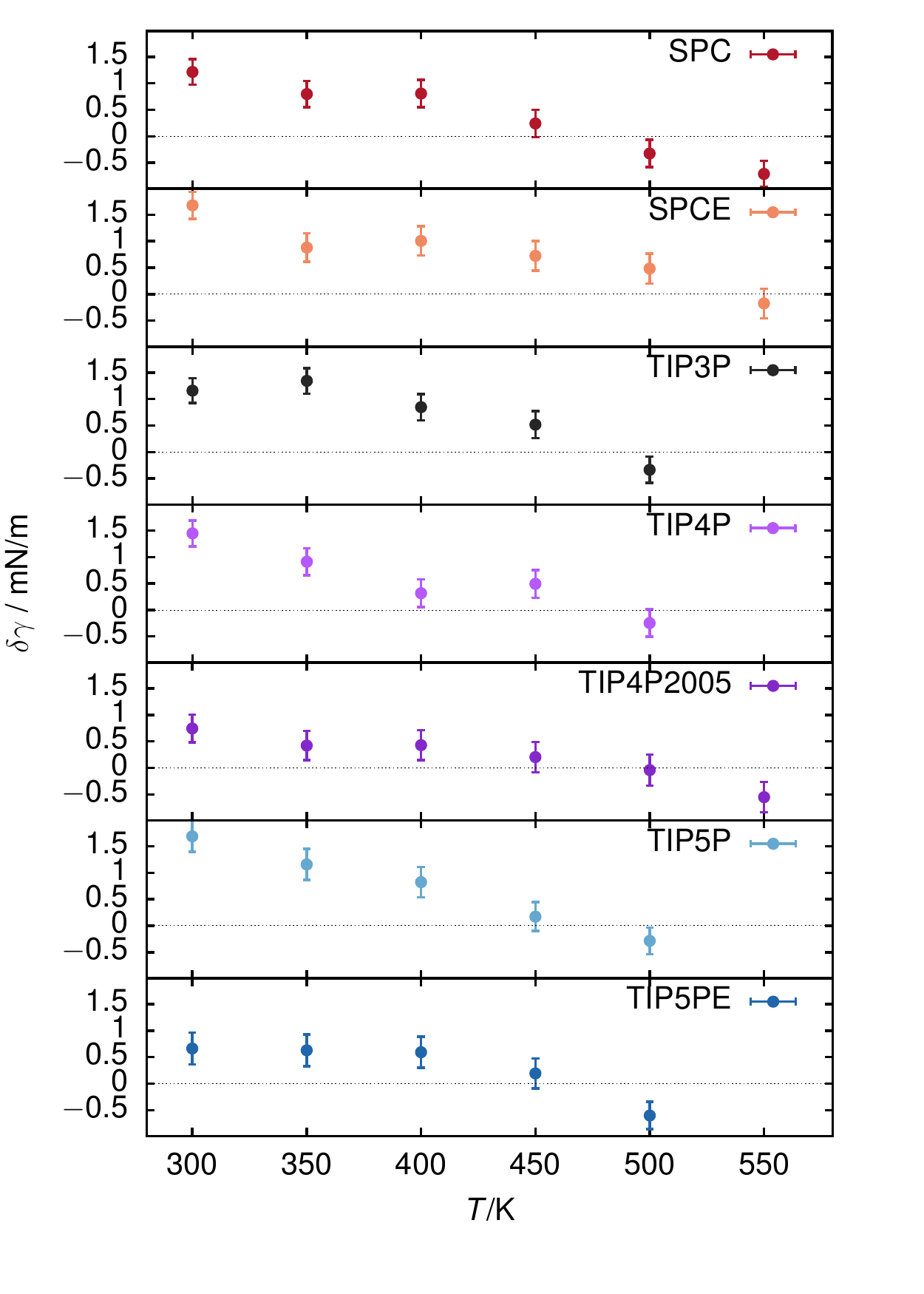}
  \caption{Surface tension difference $\delta \gamma$ between the results of the simulations with truncated dispersion forces, corrected with Eq.~(\ref{tail}) and those from the simulations with full dispersion force computed using sPME. Note that it holds also $\delta \gamma = \gamma_\mathrm{tail}-\Delta\gamma$.}
  \label{fig:diff}
\end{figure}

All water models are reproducing qualitatively the surface tension curve, which decreases towards zero as the temperature approaches the critical one, passing through an inflection point located experimentally at $T/T_c =1-b(1-\mu)/(1+\mu)=0.929$ and, for several rigid water molecules, at $T/T_c=0.70\pm0.01$ \cite{sega2014microscopic}. In Tab.~\ref{table1} we report the complete set of measured surface tensions, including those simulated with the truncated dispersion force. The long sampling times of 50~ns allowed to reach a remarkably high accuracy, with standard deviations in all cases below 0.2~mN/m. 

In Tab.~\ref{table1} we report also the value of the analytical correction estimated from the simulations with truncated dispersion forces. In order to compare the two set of results, it is more convenient to plot the difference $\delta \gamma$ between the corrected surface tension in the truncated case, and the surface tension obtained with the full-long range contributions. We have reported these values for the different water models in Fig.~\ref{fig:diff}. A positive difference $\delta \gamma$ indicates that the analytical correction is overestimating the real surface tension. In all cases, the analytical tail correction tends to overestimate the surface tension at temperatures lower than $\simeq 500$~K, where a crossover seems to occur. At first sight this seems to contradict the data on density and interfacial width, which differ more at high temperatures, but one should not forget that the tail correction has an explicit dependence on the squared difference $(\rho_L-\rho_V)^2$, which diminishes greatly the magnitude of the correction at high temperature, when the density of the two phases tend to converge to the same critical value.

\section{Conclusions}
We have calculated some interfacial properties of seven popular water models with and without inclusion of the long-range part of the dispersion forces. As the cohesion of water molecules in the liquid state is mainly determined by electrostatic forces, the binodal line is affected in a less pronounced way than in uncharged liquids, but shows, nevertheless, the clear tendency of the liquid phase to be more dense, especially at high temperature, with respect to the simulations with truncated dispersion forces. The analytical tail corrections to the surface tension  in simulations with truncated forces (with a cutoff of 1.3~nm) yield results which can differ up to 1.5~mN/m from from those obtained using the full dispersion interaction, and this can represent about 20-30\% of the analytical correction term itself at 300~K. 

The  use of mesh Ewald methods like sPME for the calculation of long-range dispersion forces  appears therefore to be advantageous even in case of liquids like water, which are dominated by the electrostatic interaction: it allows to remove the dependence of the liquid and vapor densities on the choice of the cutoff, which represents certainly an important step towards more transferable models, and it also allows to  incorporate autmatically the long-range contributions to the surface tension also for those geometries for which analytical corrections are not available.

A final note on the computational cost of the introduction of long-range dispersion forces is due: compared to the simulations with cutoff, the performance of the simulations with long-range dispersion forces decreased by 24\% for the simplest, 3 point charges models, and by 13\% for the 5 point charges models, where the overhead associated to the calculation of all other interactions is larger. The choice of the sPME parameters was, however, not aimed at maximizing the performances, but rather, to achieve accurate results.

\begin{acknowledgement}
We acknowledge financial support by the Austrian Science Foundation (FWF) within the SFB Vienna Computational Materials Laboratory (Vicom, Grant F41) and by ETN-COLLDENSE (H2020-MCSA-ITN-2014, Grant No. 642774).
Calculations were carried out on the Vienna Scientific Cluster.

\end{acknowledgement}

\bibliography{paper}

\end{document}